\documentclass[twocolumn,prc,preprintnumbers,superscriptaddress,nofootinbib]{revtex4-1}
\usepackage[utf8]{inputenc}
\usepackage{amsmath}
\usepackage{amsthm}
\usepackage{amssymb}
\usepackage{float}
\usepackage{braket}
\usepackage{graphicx}
\usepackage{url}
\usepackage{here}
\usepackage{rotating}
\usepackage{physics}
\usepackage{CJKutf8}
\usepackage{bxcjkjatype}
\usepackage[colorlinks=true, pdfstartview=FitV, linkcolor=red, citecolor=blue, urlcolor=blue]{hyperref}
\usepackage{orcidlink}
\usepackage{slashed}
\usepackage{multirow}

\newcommand{\st}[1]{{#1\!\!\!\!-}}

\begin{document}
\title{Optimal Observables for the chiral magnetic effect from Machine Learning}

\begin{abstract}
The detection of the chiral magnetic effect (CME) in relativistic heavy-ion collisions remains challenging due to substantial background contributions that obscure the expected signal. In this Letter, we present a novel machine learning approach for constructing optimized observables that significantly enhance CME detection capabilities. By parameterizing generic observables constructed from flow harmonics and optimizing them to maximize the signal-to-background ratio, we systematically develop CME-sensitive measures that outperform conventional methods. Using simulated data from the Anomalous Viscous Fluid Dynamics framework, our machine learning observables demonstrate up to 90\% higher sensitivity to CME signals compared to traditional $\gamma$ and $\delta$ correlators, while maintaining minimal background contamination. The constructed observables provide physical insight into optimal CME detection strategies, and offer a promising path forward for experimental searches of CME at RHIC and the LHC.    
\end{abstract}

\author{Yuji Hirono}
\email[]{hirono@iit.tsukuba.ac.jp}
\affiliation{Institute of Systems and Information Engineering, University of Tsukuba, Tsukuba, Ibaraki 305-8573, Japan}
\author{Kazuki Ikeda}
\email[]{kazuki.ikeda@umb.edu}
\affiliation{Department of Physics, University of Massachusetts Boston, Boston, MA 02125, USA}
\affiliation{Center for Nuclear Theory, Department of Physics and Astronomy, Stony Brook University, Stony Brook, NY 11794, USA}

\author{Dmitri E. Kharzeev}
\email[]{dmitri.kharzeev@stonybrook.edu}
\affiliation{Center for Nuclear Theory, Department of Physics and Astronomy, Stony Brook University, Stony Brook, NY 11794, USA}
\affiliation{Energy and Photon Sciences Directorate, Condensed Matter and Materials Science Division, Brookhaven National Laboratory, Upton, New York 11973-5000, USA}

\author{Ziyi Liu}
\email[]{ziyi-liu24@mails.tsinghua.edu.cn}
\affiliation{Department of Physics, Tsinghua University, Beijing 100084, China}

\author{Shuzhe Shi}
\email[]{shuzhe-shi@tsinghua.edu.cn}
\affiliation{Department of Physics, Tsinghua University, Beijing 100084, China}
\affiliation{State Key Laboratory of Low-Dimensional Quantum Physics, Tsinghua University, Beijing 100084, China}

\maketitle

\section{Introduction}
The chiral magnetic effect (CME) is a macroscopic quantum phenomenon in which an electric current is generated along an external magnetic field in systems with chirality imbalance~\cite{Fukushima:2008xe, Kharzeev:2007jp, Kharzeev:2013ffa}. As a direct manifestation of the chiral anomaly in quantum field theory, the CME has profound implications across particle physics, nuclear physics, condensed matter physics, and cosmology. In relativistic heavy-ion collisions, where intense magnetic fields ($10^{13}-10^{15}$ Tesla) are transiently created perpendicular to the reaction plane, the CME is expected to induce an electric charge separation that could provide experimental evidence for QCD topological transitions.

Despite more than a decade of experimental efforts at RHIC and the LHC, unambiguous detection of the CME in heavy-ion collisions remains elusive~\cite{STAR:2009wot, STAR:2009tro, STAR:2013zgu, STAR:2013ksd, STAR:2014uiw, STAR:2019xzd, STAR:2020gky, STAR:2021pwb, STAR:2025vhs, STAR:2025uxv, ALICE:2012nhw, ALICE:2017sss, ALICE:2020siw, CMS:2016wfo, CMS:2017lrw}. The primary challenge arises from substantial background effects that can conspire to produce similar experimental signatures~\cite{Voloshin:2004vk, Wang:2009kd, Pratt:2010gy, Bzdak:2009fc, Bzdak:2010fd, Zhao:2019hta, Li:2020dwr, Bzdak:2019pkr, Bzdak:2012ia, Choudhury:2021jwd, Christakoglou:2021nhe}. These background contributions arise from a complex interplay of local charge conservation, transverse momentum conservation, resonance decays, and collective flow phenomena, particularly elliptic flow ($v_2$), see e.g.,~\cite{Kharzeev:2024zzm, Kharzeev:2020jxw, Kharzeev:2015znc, Feng:2025yte, Wang:2024ojb, Chen:2024aom, Zhao:2019hta} for reviews. 

Conventional observables for CME detection, such as the $\gamma$ and $\delta$ correlators~\cite{Voloshin:2004vk}, measure charge-dependent azimuthal correlations but suffer from significant background contamination. These correlators cannot effectively disentangle CME-induced charge separation from background correlations, leading to inconclusive results despite extensive measurements. Recent experimental strategies, including the isobar collision program at RHIC, have attempted to address this challenge by comparing systems with different CME signals but similar backgrounds, yet definitive evidence remains pending~\cite{STAR:2021mii}.

Machine learning has emerged as a powerful approach for addressing complex data analysis challenges in high-energy nuclear physics. Recent studies have demonstrated that deep learning techniques can identify subtle patterns in heavy-ion collision data that may escape detection by traditional observables. The ability of machine learning algorithms to extract complex, non-linear relationships from high-dimensional data makes them particularly promising for the challenging task of CME detection~\cite{Aarts:2025gyp, Zhou:2023pti, Zhao:2021yjo, Guo:2025wry}.

In this Letter, we introduce a novel machine learning approach to CME detection that combines the power of optimization algorithms with physical understanding of the underlying phenomena. Rather than employing a black-box neural network, we parametrize a general class of observables constructed from flow harmonics and systematically optimize their coefficients to maximize sensitivity to CME signals while suppressing background contributions. This approach yields physically interpretable observables that significantly outperform traditional correlators in discriminating between signal and background.

Our method builds on recent theoretical advances in CME-related observables and leverages experimental insights from current measurements. We demonstrate its effectiveness using the Anomalous Viscous Fluid Dynamics (AVFD) simulation framework~\cite{Shi:2017cpu,Jiang:2016wve, Shi:2019wzi}, which provides a realistic hydrodynamic description of heavy-ion collisions incorporating both CME signals and background effects. The resulting machine learning observables offer a promising new direction for experimental CME searches that could finally provide conclusive evidence for this fundamental quantum phenomenon.

\section{method and results}

We start by defining the harmonic coefficients weighted by powers of the transverse momentum ($p_T$),
\begin{align}
    v_{n}^{(m)} \equiv \frac{\langle p_T^m\,\cos(n\,\varphi_n) \rangle}{\langle p_T^m \rangle}\,,\quad
    a_{n}^{(m)} \equiv \frac{\langle p_T^m\,\sin(n\,\varphi_n) \rangle}{\langle p_T^m \rangle}\,,
\end{align}
with all azimuthal angles $\varphi$ being defined with respect to the reaction plane. 
We then separate them into P-odd and P-even observables --- the former includes all harmonic coefficients of odd order,
\begin{align}
\begin{split}
    \{o\} \equiv 
    \left\{\left(a_{2j-1,c}^{(m)}, v_{2j-1,c}^{(m)}\right)
    \right\}_{\substack{j \in \mathbb{N},\, 0 < j \leq N/2,\\
    m \in \{0,\ldots,M\}, \\ c\in\{\pi^+, \pi^-\}}}\;,
\end{split}
\end{align}
while the latter includes the even-order ones, and in addition the inverse charge multiplicity of the event:
\begin{align}
\begin{split}
    \{e\} \equiv \left\{ \frac{1}{N_\mathrm{ch}} \right\}
    \bigcup
    \left\{\left(a_{2j,c}^{(m)}, v_{2j,c}^{(m)}\right)
    \right\}_{\substack{j \in \mathbb{N},\, 0< j \leq N/2,\\
    m \in \{0,\ldots,M\}, \\ c\in\{\pi^+, \pi^-\}}}\;.
\end{split}
\end{align}
Then we construct a potential CME observable (which should be P-even to survive the event averaging), that incorporates all terms that are linear and bilinear in P-even observables, as well as bilinear in P-odd ones:
\begin{align}
\begin{split}
O =\;&
    \sum_{j} X_{j}^{(\mathrm{L})}\, e_{j}
    + \sum_{j,j'} X^{(\mathrm{E})}_{j,j'}\, e_{j}\, e_{j'}
    + \sum_{j,j'} X^{(\mathrm{O})}_{j,j'}\, o_{j}\, o_{j'}\,.
\end{split}\label{eq:O}
\end{align}
It is straightforward to see that the conventional CME observables, the $\gamma$ and $\delta$ correlators, represent special cases of Eq.~\eqref{eq:O}.

We model heavy-ion collisions using
the Event-by-Event (EBE) Anomalous Viscous Fluid Dynamics (AVFD) simulation framework~\cite{Shi:2017cpu,Jiang:2016wve, Shi:2019wzi, An:2021wof}, which realistically simulates both the CME signal and the local charge conservation background (LCC)--the latter being a dominant source of contamination in conventional observables. We have generated $\sim 3 \times 10^{6}$ events for $30-40\%$ centrality cut of  $\mathrm{Au}+\mathrm{Au}$ collisions at $\sqrt{s_\mathrm{NN}} = 200~\mathrm{GeV}$, for the four scenarios when both CME and LCC are separately turned on and off. When activated, the CME and LCC parameters are set to $n_5/s=0.1$ and $P_\mathrm{LCC}=0.33$, respectively. The lifetime of the magnetic field is set to $0.6~\mathrm{fm}$ in the simulation. 

\begin{figure}
    \centering
    \includegraphics[width=0.5\textwidth]{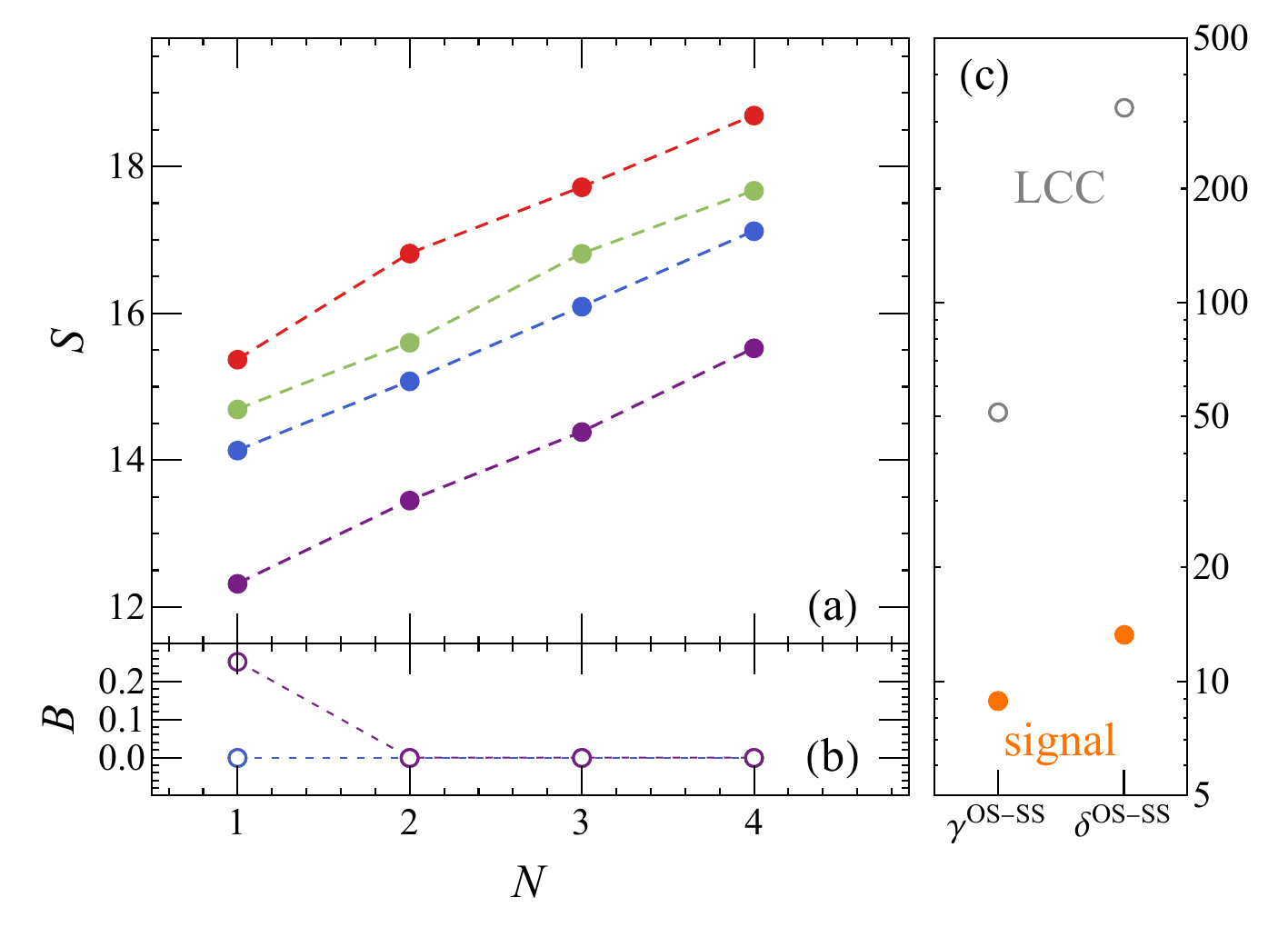}
    \caption{(a) Significance of CME signal~\eqref{eq:significance} for various cutoff in harmonic orders ($N$) and $p_T$ weights ($M$); $M$ increases from $0$ to $3$ from the bottom (purple) curve to the top (red) one; (b) The same as (a) but for a background test~\eqref{eq:background}; (c) Signal significance (orange filled) and background test (gray open) for the ``standard" $\gamma^\mathrm{OS-SS}$ (left) and $\delta^\mathrm{OS-SS}$ (right) correlators. 
    \label{fig:1}}
\end{figure}

Given a parameter set $\left\{X_{j}^{(\mathrm{L})}, X^{(\mathrm{E})}_{j,j'}, X^{(\mathrm{O})}_{j,j'}\right\}$, we calculate the means and standard deviations of $O$ for all four data sets, denoted as
\begin{align}
    O_{\mathrm{C},\mathrm{L}} = \langle O \rangle^{\mathrm{CME-on}}_{\mathrm{LCC-on}}\,,\quad
    \Delta_{\mathrm{C},\mathrm{L}} = \big(\langle O^2 \rangle^{\mathrm{CME-on}}_{\mathrm{LCC-on}} - O_{\mathrm{C},\mathrm{L}}^2\big)^{\frac{1}{2}}\,,
\end{align}
and likewise we define $O_{\mathrm{C},\st{\mathrm{L}}}$, $O_{\st{\mathrm{C}},{\mathrm{L}}}$, $O_{\st{\mathrm{C}},\st{\mathrm{L}}}$ and their corresponding standard deviations ($\Delta$'s), when one or both of CME and LCC are turned off.
Our goal is to construct an observable that maximizes sensitivity to the CME signal while minimizing responses to non-CME backgrounds, thereby addressing the primary challenge in current experimental efforts. 
To achieve this, we explicitly maximize the statistical significance of CME-induced differences while simultaneously minimizing the significance of background-induced variations, using the following loss function
\begin{align}
\begin{split}
    \mathcal{L} =\;& 
    \lambda \;\bigg(\frac{(O_{{\mathrm{C}},{\mathrm{L}}} - O_{{\mathrm{C}},\st{\mathrm{L}}})^2}{\Delta^2_{{\mathrm{C}},{\mathrm{L}}} + \Delta^2_{{\mathrm{C}},\st{\mathrm{L}}}}
+
    \frac{(O_{\st{\mathrm{C}},{\mathrm{L}}} - O_{\st{\mathrm{C}},\st{\mathrm{L}}})^2}{\Delta^2_{\st{\mathrm{C}},{\mathrm{L}}} + \Delta^2_{\st{\mathrm{C}},\st{\mathrm{L}}}}\bigg)
\\&
    -\bigg(\frac{(O_{{\mathrm{C},{\mathrm{L}}}} - O_{\st{\mathrm{C}},{\mathrm{L}}})^2}{\Delta^2_{{\mathrm{C},{\mathrm{L}}}} + \Delta^2_{\st{\mathrm{C}},{\mathrm{L}}}}
    +
    \frac{(O_{{\mathrm{C}},\st{\mathrm{L}}} - O_{\st{\mathrm{C}},\st{\mathrm{L}}})^2}{\Delta^2_{{\mathrm{C}},\st{\mathrm{L}}} + \Delta^2_{\st{\mathrm{C}},\st{\mathrm{L}}}}\bigg)\,,
\end{split}
\end{align}
where the relative factor $\lambda$ is taken as $10^4$. We then iterate the parameters $\left\{X_{j}^{(\mathrm{L})}, X^{(\mathrm{E})}_{j,j'}, X^{(\mathrm{O})}_{j,j'}\right\}$ using the gradient descent method and minimize $\mathcal{L}$. Then we compute projection of significance with $N_\mathrm{ev}=3\times10^6$ events by
\begin{align}
    S = \frac{f_c}{2}\left(
    \frac{O_{{\mathrm{C},{\mathrm{L}}}} - O_{\st{\mathrm{C}},{\mathrm{L}}}}{\sqrt{\frac{\Delta^2_{{\mathrm{C},{\mathrm{L}}}} + \Delta^2_{\st{\mathrm{C}},{\mathrm{L}}}}{N_\mathrm{ev}}}}
    +
    \frac{O_{{\mathrm{C}},\st{\mathrm{L}}} - O_{\st{\mathrm{C}},\st{\mathrm{L}}}}{\sqrt{\frac{\Delta^2_{{\mathrm{C}},\st{\mathrm{L}}} + \Delta^2_{\st{\mathrm{C}},\st{\mathrm{L}}}}{N_\mathrm{ev}}}}\right)\,.
    \label{eq:significance}
\end{align}
We have included an extra correction factor of $f_c = \frac{1}{4}$ in the estimation of signal significance to tune the assumed value of $n_5/s$\footnote{While we start from a simulation date set with axial-charge-to-entropy ratio $n_5/s=0.1$, it is found that this ratio should be chosen as $0.05$ to fit the experimental data that $f_\mathrm{CME}\equiv \gamma^\mathrm{OS-SS}_\mathrm{CME} / \gamma^\mathrm{OS-SS} \approx 0.15$~\cite{STAR:2021pwb}. The signal significance should be corrected by the factor $f_c = (0.05/0.1)^2$.}. Upon a linear response analysis of both CME signal and LCC background, we find that setting $n_5/s = 0.05$ and $P_\mathrm{LCC} = 0.33$ would fit the experimental data of both $\gamma^\mathrm{OS-SS}$ and $\delta^\mathrm{OS-SS}$.
We scan the expansion cutoffs $N=1,2,3,4$ and $M=0,1,2,3$, respectively, and show the significance of the most optimal value in Fig.~\ref{fig:1} (a).
Likewise, one may compute the background test, which is shown in Fig.~\ref{fig:1} (b), as
\begin{align}
    B = \frac{1}{2}\left(
    \frac{O_{{\mathrm{C},{\mathrm{L}}}} - O_{{\mathrm{C}},\st{\mathrm{L}}}}{\sqrt{\frac{(\Delta^2_{{\mathrm{C},{\mathrm{L}}}} + \Delta^2_{{\mathrm{C}},\st{\mathrm{L}}})}{N_\mathrm{ev}}}}
    +
    \frac{O_{\st{\mathrm{C}},{\mathrm{L}}} - O_{\st{\mathrm{C}},\st{\mathrm{L}}}}{\sqrt{\frac{(\Delta^2_{\st{\mathrm{C}},{\mathrm{L}}} + \Delta^2_{\st{\mathrm{C}},\st{\mathrm{L}}})}{N_\mathrm{ev}}}}\right)\,.
    \label{eq:background}
\end{align}

For better comparison with the conventional CME observables, $\Delta\gamma \equiv \gamma^{\mathrm{OS}-\mathrm{SS}}$ and $\Delta\delta \equiv \delta^{\mathrm{OS}-\mathrm{SS}}$, we also compute their corresponding signal significance and background test, respectively, as (with $\delta_{\Delta\gamma}$'s being the corresponding standard deviations)
\begin{align}
S_{\gamma}
    = \frac{f_c}{2}\left(
    \frac{\Delta\gamma_{{\mathrm{C},{\mathrm{L}}}} - \Delta\gamma_{\st{\mathrm{C}},{\mathrm{L}}}}{\sqrt{\frac{(\delta_{\Delta\gamma_{{\mathrm{C},{\mathrm{L}}}}})^2 + (\delta_{\Delta\gamma_{\st{\mathrm{C}},{\mathrm{L}}}})^2}{N_\mathrm{ev}}}}
    +
    \frac{\Delta\gamma_{{\mathrm{C},\st{\mathrm{L}}}} - \Delta\gamma_{\st{\mathrm{C}},\st{\mathrm{L}}}}{\sqrt{\frac{(\delta_{\Delta\gamma_{{\mathrm{C},\st{\mathrm{L}}}}})^2 + (\delta_{\Delta\gamma_{\st{\mathrm{C}},\st{\mathrm{L}}}})^2}{N_\mathrm{ev}}}}
    \right)\,,
\end{align}
and
\begin{align}
B_{\gamma}
    = \frac{1}{2}\left(
    \frac{\Delta\gamma_{{\mathrm{C},{\mathrm{L}}}} - \Delta\gamma_{{\mathrm{C}},\st{\mathrm{L}}}}{\sqrt{\frac{(\delta_{\Delta\gamma_{{\mathrm{C},{\mathrm{L}}}}})^2 + (\delta_{\Delta\gamma_{{\mathrm{C}},\st{\mathrm{L}}}})^2}{N_\mathrm{ev}}}}
    +
    \frac{\Delta\gamma_{\st{\mathrm{C}},{\mathrm{L}}} - \Delta\gamma_{\st{\mathrm{C}},\st{\mathrm{L}}}}{\sqrt{\frac{(\delta_{\Delta\gamma_{\st{\mathrm{C}},{\mathrm{L}}}})^2 + (\delta_{\Delta\gamma_{\st{\mathrm{C}},\st{\mathrm{L}}}})^2}{N_\mathrm{ev}}}}
    \right)\,.
\end{align}
Likewise for $S_{\delta}$ and $B_{\delta}$. They are shown in Fig.~\ref{fig:1} (c).

With a long magnetic field lifetime and a relatively large initial chirality imbalance, we observe that both $\gamma^{\mathrm{OS}-\mathrm{SS}}$ and $\delta^{\mathrm{OS}-\mathrm{SS}}$ correlators are sensitive to  
CME signal, with statistical significance of $8.9$ and $13.3\ \sigma$, respectively. However, they are even more influenced by the LCC backgrounds (background-to-error ratios are $52$ and $327$, respectively), which makes signal identification extremely challenging. 

Our optimized observables demonstrate substantial improvements even at the lowest level of complexity.
With terms up to $N=1$ and $M=0$, the constructed observable is essentially an optimized combination of $\gamma^{\mathrm{OS}}$, $\gamma^{\mathrm{SS}}$, $\delta^{\mathrm{OS}}$, $\delta^{\mathrm{SS}}$, and inverse charge multiplicity (dependence of the elliptic flow might enter in an implicit manner). While its sensitivity to the CME is $12.3$, which is comparable to those in $\gamma^{\mathrm{OS}-\mathrm{SS}}$ and $\delta^{\mathrm{OS}-\mathrm{SS}}$, it dramatically reduces background influence to only $\sim 0.2$.

By increasing $N$ and $M$, we can systematically enhance the signal significance, ultimately reaching $18.7\ \sigma$ with $N=4$ and $M=3$. Notably, the influence of background on the observable is consistent with zero for $N\geq2$, i.e. when the elliptic flow is included. As expected from the background models, the non-CME contributions to $\gamma^{\mathrm{OS}-\mathrm{SS}}$ and $\delta^{\mathrm{OS}-\mathrm{SS}}$ are proportional to $v_2/N_\mathrm{ch}$ and $1/N_\mathrm{ch}$, respectively. This means that, including terms with $N=2$, the optimization procedure can eliminate this background.
For illustration, the explicit form of the optimized observable when including terms up to $N=2$
without $p_T$ weighting ($M=0$) is:
\begin{align}
\begin{split}
    O =\;& 0.517\, ((a_{1}^+)^2 + (a_{1}^-)^2) 
    - 0.048\, ((v_{1}^+)^2 + (v_{1}^-)^2) 
\\&
    + 0.618\,(v_{1}^+ v_{1}^- - a_{1}^+ a_{1}^-) \,
\\&
    + \frac{0.104}{N_\mathrm{ch}} 
    + \frac{6.60\times10^{-5}}{N_\mathrm{ch}^2}
    + 1.25\times10^{-2}\frac{v_2^++v_2^-}{N_\mathrm{ch}}
\\&
    + 0.0166\,v_2^+ - 0.099\,v_2^-
\\&
    - 0.0199\,(v_2^+)^2 - 0.190\,(v_2^-)^2
    - 0.0867\,v_2^+\,v_2^-
\\&
    + 0.0199\,(a_2^+)^2 - 0.154\,(a_2^-)^2
    + 0.0713\,a_2^+\,a_2^-
\\&
    - 0.0504\,v_2^+\,a_2^-
    + 0.0866\,a_2^+\,v_2^-
    \,.
\end{split}\label{eq:obs}
\end{align}

\section{Conclusion and Discussion}

In this Letter, we have developed a novel machine learning approach to construct optimized observables for detecting the CME in heavy-ion collisions. Our method creates a (P-even) superposition of terms linear and bilinear in harmonic coefficients weighted by powers of the transverse momentum. By systematically optimizing these superposition coefficients, we have achieved three significant advances: (1) observables that effectively suppress non-CME background contributions to near-zero levels, (2) up to 110\% greater sensitivity to the CME signal compared to conventional $\gamma$ correlators, relative to their statistical errors, and (3) a framework that enables systematic improvement in signal significance through the inclusion of higher-order harmonics and transverse momentum weighting. Importantly, this approach is readily extensible to additional parameters such as rapidity, centrality, particle species, or any other experimentally accessible quantities, providing further opportunities to optimize CME detection in experimental data.

While the specific coefficients in our optimized observable~\eqref{eq:obs} depend on the AVFD model parameters used in this study, the underlying methodology itself provides immediate benefits for experimental applications.  
When applied to experimental data, our approach may offer improved performance compared to conventional correlators even if the actual background mechanisms differ somewhat from our simulations.  
Furthermore, the same machine learning approach can be adopted to find an ``ideal'' CME observable $O_\mathrm{CME}$ directly from experimental data.  
This can be achieved by minimizing the observable response in ultra-high-energy collisions ($\sqrt{s_\mathrm{NN}} \geq 5.02~\mathrm{TeV}$), where the CME is expected to be negligible due to the rapid decay of the magnetic field before quark production, while maximizing the signal difference between $\mathrm{Ru}+\mathrm{Ru}$ and $\mathrm{Zr}+\mathrm{Zr}$ collisions. 
That is, one may construct $O_\mathrm{CME}$ defined Eq.~\eqref{eq:O} with the $X$ coefficients being independent of collision system, energy, and centrality. These coefficients are then optimized by minimizing the loss function:
\begin{align}
\begin{split}
    \mathcal{L} =\;& 
    \lambda \left(\sum_{c\in\{\text{centrality bins}\}}\frac{O_{\mathrm{Null},c}^2}{\Delta^2_{\mathrm{Null},c}}
\right)
\\&
    -\left(\sum_{c\in\{\text{centrality bins}\}}\frac{(O_{\mathrm{Ru},c} - O_{\mathrm{Zr},c})^2}{\Delta^2_{\mathrm{Ru},c} + \Delta^2_{\mathrm{Zr},c}}
    \right).
\end{split}
\end{align}
A simple but evident criterion to test whether one has found the ``ideal'' CME observable exists: a genuine CME signal should follow the expected scaling relation $O_\mathrm{CME}^{\mathrm{Ru}+\mathrm{Ru}} \approx 1.2\, O_\mathrm{CME}^{\mathrm{Zr}+\mathrm{Zr}}$, which arises from the scaling of the squared magnetic field strength.

\section*{Acknowledgment}
The authors thank Jinhui Chen, Jinfeng Liao, Diyu Shen, and Lingxiao Wang for helpful discussions.
Z.~L. and S.~S. are supported by National Key Research and Development Program of China under Contract No. 2024YFA1610700 and by Tsinghua University under grants No. 043-531205006 and No. 043-53330500.
Y.\,H. is supported in part by JSPS KAKENHI Grant Numbers JP22H05111, JP22H05118, JP24K23186, and by JST, PRESTO Grant Number JPMJPR24K8.
D. K. is supported by the U.S. Department of Energy, Office of Science, Office of Nuclear Physics, Grants No. DE-FG88ER41450 and DE-SC0012704 and by the U.S. Department of Energy, Office of Science, National Quantum Information Science Research Centers, Co-design Center for Quantum Advantage (C2QA) under Contract No.DE-SC0012704.

\section*{Data Availability}
The data that support the findings of this article are openly available~\cite{shi_2025_17586247}.

\bibliography{ref}

\end{document}